\documentclass[letterpaper,aps,pre,floats,twocolumn,superscriptaddress,showpacs]{revtex4}
\usepackage{graphicx}
\usepackage{epsfig}
\begin{document}
\title{Nonperiodic echoes from mushroom billiard hats}

\author{B.~Dietz}
\affiliation{Institut f\"{u}r Kernphysik, Technische
Universit\"{a}t Darmstadt, D-64289 Darmstadt, Germany}

\author{T.~Friedrich}
\affiliation{Institut f\"{u}r Kernphysik, Technische
Universit\"{a}t Darmstadt, D-64289 Darmstadt, Germany}

\author{M.~Miski-Oglu}
\affiliation{Institut f\"{u}r Kernphysik, Technische
Universit\"{a}t Darmstadt, D-64289 Darmstadt, Germany}

\author{A.~Richter}
\affiliation{Institut f\"{u}r Kernphysik, Technische
Universit\"{a}t Darmstadt, D-64289 Darmstadt, Germany}

\author{T.~H.~Seligman}
\affiliation{Centro de Ciencias F\'{\i}sicas, UNAM, Mexico,
Mexico}
\affiliation{Centro Internacional de Ciencias, Cuernavaca,
Mexico}

\author{K.~Zapfe}
\affiliation{Centro Internacional de Ciencias, Cuernavaca, Mexico}

\date{\today}
\begin{abstract}
Mushroom billiards have the remarkable property to show one or
more clear cut integrable islands in one or several chaotic seas,
without any fractal boundaries. The islands correspond to orbits
confined to the hats of the mushrooms, which they share with the
chaotic orbits. It is thus interesting to ask how long a chaotic
orbit will remain in the hat before returning to the stem. This
question is equivalent to the inquiry about delay times for
scattering from the hat of the mushroom into an opening where the
stem should be. For fixed angular momentum we find that no more
than three different delay times are possible. This induces
striking nonperiodic structures in the delay times that may be of
importance for mesoscopic devices and should be accessible to
microwave experiments.
\end{abstract}

\maketitle

\section{Introduction}
Two-dimensional billiards play a central role in the development
of chaos theory ever since the early work of Sinai \cite{sinai}.
In physics they have acquired increasing importance as they are
seen to emulate properties of systems as different as quantum dots
\cite{qdots} or planetary rings \cite{benet}. Particularly quantum
or wave realizations of such objects have become popular, since
flat microwave cavities, so-called microwave billiards, are used
by experimentalists at different laboratories \cite{smilansky,
stoeckmann, richter, sirko, anlage, sridhar, legrand}, and these
experiments mimic some properties of mesoscopic devices. From a
mathematical point of view one of the advantages of billiards is
that, in many instances, chaotic properties can be proven for
cases of complete chaos
\cite{sinai,bunimovich1,threedisk,threehills} and more recently
even for mixed systems \cite{bunimovich2,jung}.
\par
Microwave billiards serve as an analog system for the experimental
study of the wave behavior of the corresponding classical
billiard, and permit a direct test of hypotheses proposed
concerning connections between the quantum and the corresponding
classical dynamic  \cite{Stockmann,IMAPRO}. In particular the
spectral behavior and properties of the wave functions of quantum
systems, whose classical analogue is chaotic \cite{Casati,
Bohigas, Berry, Leyvraz, Sieber, Braun-Haake}, integrable
{\cite{Gutzwiller, Berry-Tabor}} or intermediate
\cite{SVZ,Berry-Robnik,SV-JPA} can be investigated experimentally
with such systems. In scattering systems similar matters have been
discussed mainly for systems that are chaotic and hyperbolic
\cite{threedisk,Jung-Dietz}, or mixed
\cite{Ketzmerick,Jung-Seligman,carlosnjp,mejia,friedrich}. The
role of parabolic manifolds has also received considerable
attention \cite{richter,Dietz-Lombardi}.
\par
\begin{figure}
\begin{center}
\epsfig{figure=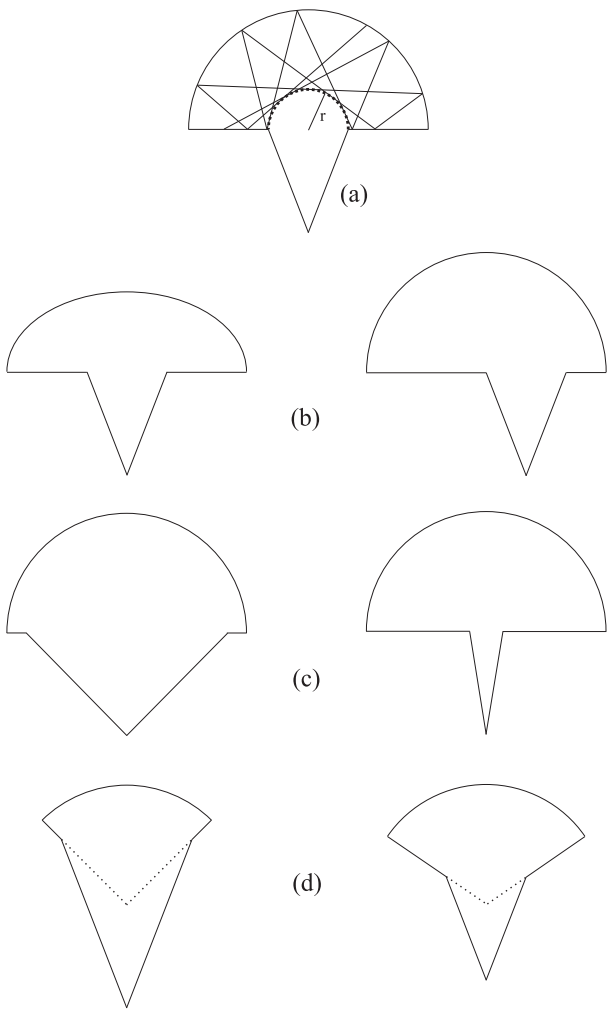,width=8cm,angle=0} \caption{\label{fig_1}
Different shapes of mushroom hats. Circular mushroom billiards
consist of a hat with a circular boundary and a stem, which has
been chosen in triangular form for all mushroom billiards here.
Stable orbits in a circular mushroom hat never enter the stem.
They form a caustic in the hat, and the radius of the smallest
possible caustic is one-half the width of the stem, as indicated
in (a). Other shapes of mushroom billiards may display an elliptic
hat or a shifted stem (b), very wide or narrow stems (c) or hats
which are a circle sector of an angle smaller than 180$^\circ$
(d), in this case $90^\circ$ and $180^\circ\times (\sqrt5-1)/2$.}
\end{center}
\end{figure}

Mixed systems typically have the property, that in some border
region integrable and chaotic areas intermingle as a fractal. Such
a scenario is well illustrated by the twist map \cite{twistmap}.
It may help to detect some characteristics of the dynamics in
certain cases \cite{friedrich}. Bunimovich \cite{bunimovich2} has
proposed a family of mushroom billiards with mixed phase space
which have the unusual property that chaotic areas and integrable
ones are not separated by a fractal set of integrable islands
extending into the chaotic sea. Such billiards are characterized
by circular or elliptic \textit{hats} connected to a \textit{stem}
or stems composed of straight walls; we shall concentrate on the
case with only one hat and one stem. Typical examples are shown in
Fig.~\ref{fig_1}. The stem pertains entirely to the chaotic area,
while the hat houses both chaotic and integrable trajectories.
Several studies on the classical dynamics in mushroom billiards
have already been presented \cite{altmann,mushroomworks}.
\par
It is clear, that the relevant properties of such a billiard are
determined by the hat of the mushroom, and in view of the
successful theoretical \cite{mejia} and experimental
\cite{friedrich} analysis of phase space structures in the context
of scattering echoes we wish in the present paper to analyze the
hat of the mushroom alone, viewed as a scattering system with an
opening where the stem used to be; yet to establish contact with
other work, we shall also keep the possibility of a stem in mind.
In the spirit of scattering echoes we ask how long a particle
stays in the hat, if injected from the opening of the billiard or
from the stem. Note that in the case of a semicircle mushroom hat
with a symmetric stem, the limit of the area of stable bounded
orbits is a caustic, which forms a circle segment connecting the
points where the stem starts, as seen in the mushroom billiard in
Fig.~\ref{fig_1}(a). The usual Smale horseshoe construction
\cite{jung} fails, because there is no hyperbolic periodic orbit
along this line. We have therefore numerically investigated the
distribution of the classical delay times of such a billiard and
found a surprising selectivity in the allowed delay times.
\par
As the angular momentum is a constant of motion in the scattering
process as long as a particle stays within the hat we shall first
consider fixed angular momenta and we find that generically only
three delay times occur. On a subset of measure zero, one
and two delay times are also possible; a larger set of different
delay times is outright forbidden. This result is intimately
related to old results on the circle map \cite{slater} and allows
to understand the above mentioned selectivity.

\section{Definitions and preliminary results}
The circular hats of mushrooms are characterized by two
quantities: The position and size of the hole and the angle
between the straight walls that constitute the {\it underside}\ of
the hat. In Fig.~\ref{fig_1}~we show various hats of mushroom
billiards with different properties: circular mushroom hats [(a)
and (c)], mushroom billiards with a shifted stem or an elliptic
hat (b) and mushroom billiards whose hat is a circle sector of an
angle different from $180^\circ$ (d). The stem is always chosen in
triangular form to avoid the trivial parabolic manifold of
so-called "bouncing ball" orbits between parallel walls of a
rectangular stem.
\begin{figure}
\begin{center}
\epsfig{figure=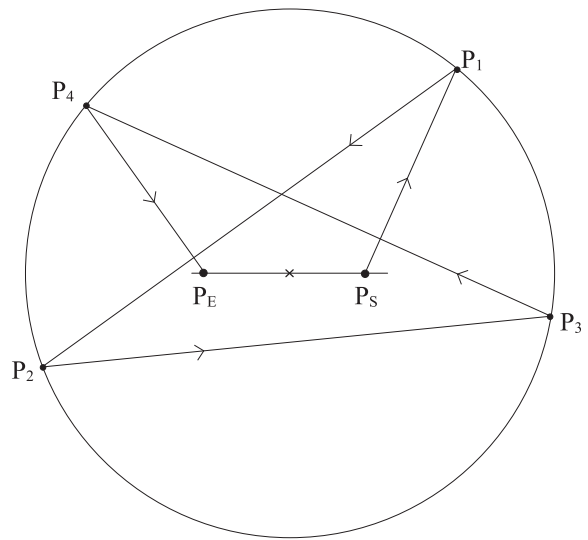,width=8cm,angle=0} \caption{\label{fig_2}
The $\Theta$~billiard is a circle billiard with a straight line of
length $2r$ along a diameter of the circle of length $2R$, which
defines the starting and ending points of the particle orbits
taken into consideration. In a semicircle mushroom hat with the
opening located at the position of the straight line these orbits
correspond to those of particles entering and leaving it. In this
sketch, a particle starts from the straight line at $P_S$,
undergoes four reflections with the boundary at $P_1$, $P_2$,
$P_3$, and $P_4$ before it finally ends on the straight line at
$P_E$.}
\end{center}
\end{figure}
\par
In Fig.~\ref{fig_1}(c) we show $180^\circ$-mushroom billiards with
a wide and a narrow stem, respectively. Here a narrow stem is one
narrow enough, that orbits which are triangular in the full circle
remain within the hat irrespective of the orientation of the
triangle. This implies, that all polygons, which do not form stars
with intersecting line segments are confined to the hat. Moreover,
the delay time, i.e., the time a particle entering the mushroom
hat stays there, is approximately determined by the number of its
bounces off the circle boundary. This is clearly not the case for
a wide stem, where some polygons are no longer confined to the
hat. In the following we will restrict ourselves to mushroom
billiards with a narrow stem and count bounces of particles
entering and leaving the mushroom hat, instead of determining
delay times.
\par
We also show mushroom billiards where the two straight walls
constituting the underside of the hat have angles smaller than
$180^\circ$ in Fig.~\ref{fig_1}(d). Note that it is essential that
the two straight walls are parts of radii of the circle, i.e.,
form an angle of $90^\circ$ with the tangent to the circle
boundary at their intersection. Otherwise we lose the properties
of the mushroom billiard. If the angle is larger than $90^\circ$,
such that the extensions of the two straight walls meet in a point
below the center of the circle, the system becomes ergodic and
chaotic \cite{bunimovich3}. For an angle smaller than $90^\circ$
we get a more complex coexistence of integrable and chaotic
dynamics.
\par
In the following we first will consider open $180^\circ$-mushroom
billiards with a semicircular hat and a symmetric stem. We are
interested in the number of bounces a particle entering a mushroom
hat through the opening experiences at the circle boundary before
exiting again. Bounces with the straight part of the boundary are
not counted. Whenever the particle hits it, we may use the
principle of mirror images \cite{princ}, i.e., reflect the
semicircle across its straight boundary, thereby obtaining a
complete circle and continue the orbit into its lower half.
Accordingly, we may introduce a convenient alternative system, the
so-called $\Theta$~billiard \cite{altmann} shown in
Fig.~\ref{fig_2}, which consists of a circle billiard, whose
radius $R$ equals that of the mushroom hat, and a straight line of
length $2r$ where the opening of the corresponding semicircle
mushroom hat is. The only purpose of the straight line is, that it
defines the starting and the endpoints of those particle orbits we
are interested in. In the mushroom billiard these correspond to
particles entering and leaving the semicircle hat. Accordingly, in
order to compute delay times, we consider particle orbits starting
on the straight line in the $\Theta$~billiard and count the number
of bounces of the particle with the circle boundary, until it
again reaches the straight line. We shall see that the
$\Theta$~billiard can be readily generalized to describe arbitrary
mushroom hats. While the $\Theta$~billiard is convenient for
mathematical purposes physically it seems of little direct use.
\par
As already noted above and indicated in Fig.~\ref{fig_1}(a), for
every particle orbit in the circle billiard the line segments
connecting subsequent reflection points form a caustic of circular
shape around the center of the billiard. The radius of the caustic
equals the distance of these line segments from the circle center.
Therefore it gives the angular momentum of the particle where we
set the absolute value of the momentum to unity, as it is a
conserved quantity. If the angular momentum of the particle is
large enough, the corresponding orbit will never hit the straight
line in the $\Theta$~billiard. Such orbits belong to the
integrable part of the phase space of the mushroom billiard. If
they are periodic they may be polygons or stars. If they are
aperiodic they will correspond to slowly rotating stars or
polygons, i.e., they would be periodic in some rotating frame of
reference. The limiting angular momentum corresponds to the radius
of that caustic, which just touches one or both ends of the
straight line. Any aperiodic orbit with angular momentum smaller
than the limiting value will eventually reach the straight line
after some number of bounces. For periodic orbits we can have the
particular case, where a star or polygon will intersect the
straight line in certain orientations and not in others. The
orientations that do not lead to intersections will yield a
parabolic manifold in the chaotic sea as explained in Ref.~
\cite{altmann}. This paper focuses on orbits that do cross the
straight line.
\par
We now come to a really surprising result. Using a simple
reflection program, which computes orbits of a particle in a
billiard based on the law of specular reflection, we counted the
bounces off the circle boundary a particle starting from the
straight line with randomly chosen initial conditions experiences
until it reaches the straight line again. We found a strange
selectivity in these numbers, which resembled a generalized
Fibonacci sequence. Indeed for $r=R/3$ the first observed numbers
of bounces are 1,~4,~5,~9,~14,~23,~37. The next number in this
sequence, 51, does not continue the Fibonacci series, but is the
sum of 37 and 14. Similar results have been found for other ratios
$r/R$. We shall call the possible numbers of bounces.
\par
The purpose of this paper is to understand the observed
selectivity. For this we will introduce a map, which generates the
dynamics in the mushroom hat in an efficient way and then shall
proceed in two steps utilizing this map. First we shall analyze
the problem for a fixed angular momentum, considering that angular
momentum is a conserved quantity both in the open mushroom hat and
in the $\Theta$~billiard. We shall obtain the surprising result
that typically only three magic numbers termed a \textit{magic
triplet} are possible for each angular momentum. To understand the
entire delay time structure, we then analyze how these triplets
evolve when changing the angular momentum. Based on the map, we
developed a very efficient algorithm for the computation of the
magic numbers.

\section{Magic numbers for fixed angular momentum}
\begin{figure}
\begin{center}
\epsfig{figure=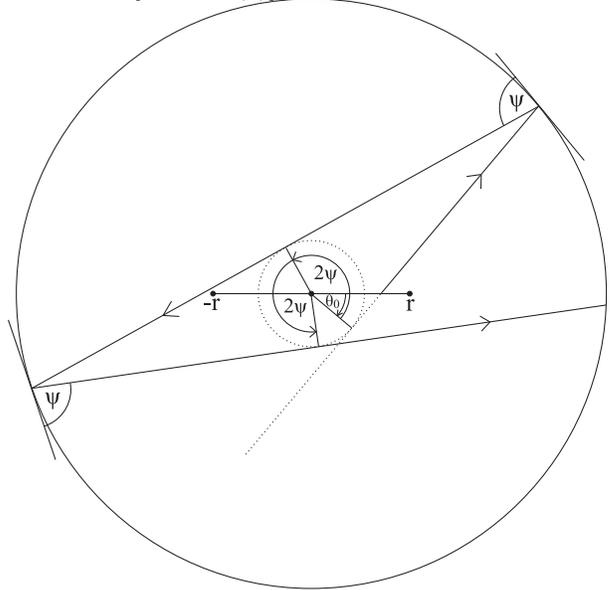,width=8cm,angle=0} \caption{\label{fig_3}
A $\Theta$~billiard with the straight line chosen symmetric with
respect to the center of the circle. Initial conditions for a
particle orbit in the $\Theta$~billiard are the angular momentum
$M$, it i.e., the radius of its caustic (dotted line circle), and
the orientation $\theta_0$ of its first line segment. The
orientation of each line segment of a particle orbit is defined in
terms of the total angle covered by the vector pointing from the
center of the circle to the point of contact of the line segment
with the caustic during its rotation while the particle
propagates. In this example the orientation $\theta_0$ of the
initial line segment is negative and determined by extending the
latter back beyond its point of contact with the caustic as
indicated in the figure by the dotted line. The angle $\psi$
between the reflected particle orbit and the tangent to the circle
boundary at the point of reflection is a constant. At each
reflection the orientation angle increases by $2\psi$.}
\end{center}
\end{figure}

Let the radius of the $\Theta$~billiard be $R$ and the straight
line of length $2r$ be oriented symmetrically along the horizontal
axis. A particle orbit is characterized by the particle's angular
momentum $M$, which is just the radius of its caustic. Orbits with
angular momentum $M>r$ never intersect the straight line, whereas
orbits with $M<r$ intersect it (up to a set of marginally unstable
orbits of measure zero as stated above) at some time and therefore
correspond to chaotic trajectories in the mushroom billiard. The
angular momentum can be expressed in terms of the angle between
the orbit of the reflected particle and the tangent to the circle
boundary at the point of impact, which also is a constant for each
particle orbit; we will call this angle $\psi$, where
$0<\psi\leq\pi /2$ (see Fig.~\ref{fig_3}). Then, the radius of the
caustic, i.e., the angular momentum $M$ equals
\begin{equation}
M=R\cos\psi\,. \label{psidef}
\end{equation}
A particle orbit is also characterized by its initial orientation.
Generally, the orientation of the particle orbit after $n$
reflections at the circle boundary may be given in terms of the
orbit's angle with the positive horizontal axis. Equivalently, its
orientation may be defined in terms of the total angle covered by
the vector pointing from the circle center to the point of contact
between the caustic and the orbit during its rotation while the
particle is being reflected at the circle boundary. For our
purposes it is more convenient to define it in terms of the
latter. We denote this angle by $\theta_n$ for the $n$th line
segment of a particle orbit, which connects the $n$th point of
impact with the $(n+1)$th. At each reflection the angle increases
by $2\psi$ (see Fig.~\ref{fig_3}). Then, for a particle orbit
starting from the straight line with an initial orientation
$\theta_0$ measured with respect to the positive horizontal axis
and an initial angular momentum $M$, which defines $\psi$,
$\theta_n$ reads
\begin{equation}
\theta_n=\theta_0+n2\psi. \label{Thetan}\end{equation} Note, that
the angle $\theta_n$ is not restricted to the interval $[0,2\pi)$,
as it also contains the information on the number of times the
particle travelled around the circle center. In conclusion, each
particle orbit is completely characterized by the initial values
$(\psi,\theta_0)$, the radius of the caustic [Eq.~(\ref{psidef})]
and the orientation of each of its line segments
[Eq.~(\ref{Thetan})].
\par
Due to the symmetry of the system it is sufficient to deal with
orbits proceeding from the straight line into the right upper
quarter of the $\Theta$~billiard. For a given initial orientation
$\theta_0$ the orbit starts on the horizontal line at a distance
$M/\cos\theta_0$ from the circle center (see Fig.~\ref{fig_3}). As
we are only interested in orbits starting from the straight line,
which reaches from $-r$ to $r$, we obtain as a necessary condition
for the initial angle $\theta_0$
\begin{equation}
-\chi\le \theta_0 \le \chi\label{winkelbereich}
\end{equation}
with the notation
\begin{equation}
\chi=\arccos(M/r)\label{chidef},
\end{equation}
where with Eq.~(\ref{psidef}),
\begin{equation}
0<\chi\le\psi\label{winkelbereichchi}.
\end{equation}
For an angular momentum $M$ and length $2r$ of the straight line,
$\pm\chi$ in Eq.~(\ref{chidef}) correspond to the initial
orientation of those particle orbits which start at the endpoints
of the straight line. For a particle starting with an initial
orientation $\theta_0$ the $n$th line segment of its orbit
intersects the horizontal line at $M/\cos\theta_n$, if measured
with respect to the circle center. Hence, it crosses the straight
line, if $-r\leq M/\cos\theta_n\leq r$, i.e., if [with
Eq.~(\ref{chidef})]
\begin{equation}
\theta_n\,\mathrm{mod}\,2\pi\in
[0;\chi]\cup[\pi-\chi;\pi+\chi]\cup[2\pi-\chi;2\pi]\;,
\label{map}
\end{equation}
or equivalently if there is an integer number $m$ such that
\begin{equation}
m\pi-\chi\leq\theta_n\leq m\pi+\chi\ ,\ m=0,1,\cdots\ ,
\label{ineq}
\end{equation}
that is, if the distance between $\theta_n$ and an integer
multiple of $\pi$ is smaller than $\chi$. In order to determine
all possible numbers of bounces that particles starting from the
straight line perform until they reach it again, we must evaluate
this inequality for all allowed initial orientations $\theta_0$
[see Eq.~(\ref{winkelbereich})] and resulting orientations
$\theta_n$. With Eq.~(\ref{winkelbereich}) and Eq.~(\ref{Thetan})
the latter take values from intervals of length $2\chi$ around
integer multiples of $2\psi$,
\begin{equation}
n2\psi-\chi\leq\theta_n\leq n2\psi +\chi\ ,\ n=0,1,\cdots .
\label{ineqth}
\end{equation}
While this inequality defines the range of possible values for the
orientations $\theta_n$, that in Eq.~(\ref{ineq}) gives the
condition for the intersection of a particle orbit with the
straight line after $n$ reflections. Whenever one of the intervals
defined in Eq.~(\ref{ineqth}) has common values with one of those
given in Eq.~(\ref{ineq}), a part of all possible orbits reaches
the straight line. Thus, the only information we need for the
computation of the magic numbers are the orientations of the line
segments of a particle orbit. This procedure for the computation
of the magic numbers is fast and much more efficient than the
reflection program. \begin{widetext} \onecolumngrid
\begin{figure}
\begin{center}
\epsfig{figure=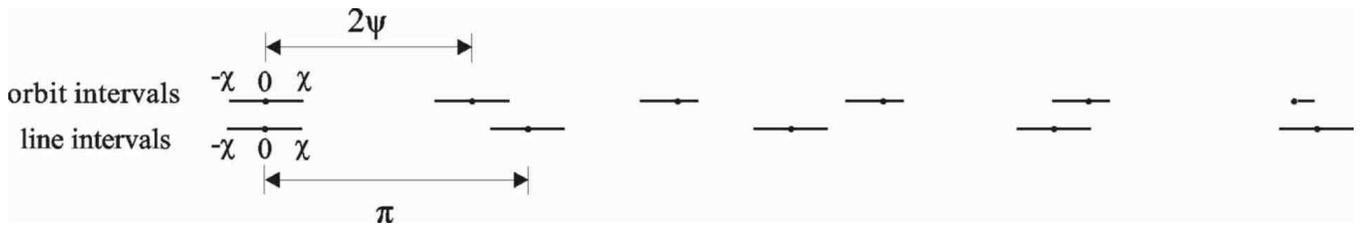,width=\linewidth,angle=0}
\caption{\label{fig_4} Illustrative representation of particle
orbits starting from the straight line in the $\Theta$~billiard in
terms of the orientations of their line segments. The orientations
of the ensemble of all particle orbits starting from the straight
line are restricted to the interval $[-\chi ,\chi]$ [see
Eq.~(\ref{winkelbereich}) in the main text]. Their evolution is
illustrated by shifting this interval by $2\psi$  at each
reflection of the particle with the billiard boundary, thereby
obtaining the range of values of the orientations of successive
line segments of the particle orbits [see Eq.~(\ref{ineqth})].
Thus, a string of "orbit intervals" is obtained. The straight line
is represented by a string of "line intervals" of size $2\chi$
around each multiple of $\pi$ on the angle axis. When an overlap
of a line and an orbit interval occurs, the overlapping part of
the orbit interval corresponds to angles $\theta_n$, which fulfill
the inequality Eq.~(\ref{ineq}), that is, to initial angles
$\theta_0$, for which the particle orbit ends after $n$
reflections. Accordingly, the overlapping part is cut off the
orbit interval in the subsequent iterations.}
\end{center}
\end{figure}
\end{widetext}
\par
For a given angular momentum $M$, which defines the angle $\psi$,
and length of the straight line $2r$ giving the angle $\chi$, an
illustrative graphical representation of this inequality is
obtained as follows: All possible initial orientations $\theta_0$
of the particle orbits we are interested in take a value from an
interval of length $2\chi$ situated symmetrically around 0 [see
Eq.~(\ref{winkelbereich})]. According to Eq.~(\ref{ineqth}) the
orientation $\theta_n$ of the $n$th line segment of a particle
orbit then takes a value from an interval of length $2\chi$
shifted by $n2\psi$ with respect to the initial interval and by
$2\psi$ with respect to that of the preceding line segment,
$\theta_{n-1}$. Hence, the orientations of the line segments of
all those particle orbits, which start from the straight line,
resume values from a string of angle intervals of length $2\chi$
situated at distances of $2\psi$ along the angle axis as depicted
in Fig.~\ref{fig_4}. We call these intervals orbit intervals.
Next, let us draw a string of intervals of length $2\chi$ at
distances $\pi$ along the angle axis below the string of orbit
intervals; we will call these intervals line intervals. While the
orbit intervals refer to the inequality Eq.~(\ref{ineqth}) the
latter refer to the inequality Eq.~(\ref{ineq}). The inequality
Eq.~(\ref{ineq}) is fulfilled for a part of the interval of
allowed initial angles, whenever one of the orbit and one of the
line intervals have common values, i.e., partially overlap. As we
are only interested in orbits which start from the straight line,
the first orbit interval completely overlaps with the first line
interval.
\par
In the example shown in Fig.~\ref{fig_4}, an orbit interval
overlaps with a line interval after the first shift of the orbit
interval. Hence, a part of the incoming particles reaches the
straight line after one bounce with the circle boundary and the
corresponding orbits end there. Accordingly, the orbit interval is
shortened by this part. After four reflections, another part of
all possible particle orbits reaches the straight line, i.e., the
orbit interval is further shortened. The remaining part of the
orbit interval overlaps with a line interval after five
reflections. Thus, in this example the magic triplet consists of
the numbers 1, 4, and 5.
\par
The Eqs.~(\ref{Thetan}) and (\ref{map}) are related to old results
on the so-called circle map \cite{slater}. On the basis of these
results it can be shown that, for a fixed angular momentum, there
are at most three magic numbers, and that if there are three magic
numbers the largest one is the sum of the two smaller ones. We
found an alternative proof of this result which in fact
generalizes the translation of orbit and line intervals as
depicted in Fig.~\ref{fig_4}. Using this approach we also obtained
the following results: There is a subset of measure zero, where we
have only two magic numbers. Further, in the case of the parabolic
manifolds described in Ref. \cite{altmann} only one magic number
is finite, whereas the others are infinite and correspond to
marginally unstable orbits which never leave the mushroom hat.
Finally, in the case of angular momentum zero, all particle orbits
intersect the straight line after a single bounce. We thus have
exhausted all possibilities. We give in Table~\ref{tab_1} the
magic triplets for different angular momenta for $r=R/3$ used in
Fig.~\ref{fig_1}(a), and we readily recognize the numbers therein.
\begin{widetext}
\onecolumngrid
\begin{table}[!t]
\caption{\label{tab_1} Magic triplets for different values of
angular momentum $M$ for a circular mushroom hat with radius $R$
and half-opening $r=R/3$.}
\begin{centering}
\vspace{0.5cm} \onecolumngrid
\begin{tabular}{c||c|c|c|c|c|c|c|c|c}
\hline \hline
$M$&0.1000&0.5000&0.9000&0.9500&0.9800&0.9950&0.9980&0.9990&0.9992\\
\hline \hline
$n_1$&1&1&1&4&5&9&14&14&37\\
$n_2$&4&4&4&5&9&14&23&37&51\\
$n_3$&5&5&5&9&14&23&37&51&88\\
\hline \hline
\end{tabular}
\end{centering}
\end{table}
\end{widetext}
\section{Magic numbers as a function of angular momentum}
\begin{figure}[b]
\begin{center}
\epsfig{figure=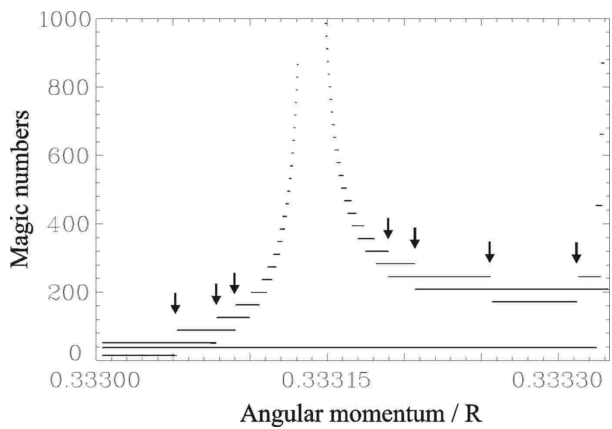,width=8cm,angle=0} \caption{\label{fig_5}
Magic numbers plotted as a function of angular momentum for a
semicircle mushroom hat with radius $R$ and a half-opening $r=R/3$
chosen symmetrically with respect to the circle center. The
angular momentum is given in units of $R$ as the absolute momentum
is set to unity. One observes lines where a magic number remains
constant over a certain range of angular momentum values, and
sudden discontinuities (arrows), where one member of the triplet
changes its value. Note further a singularity, where two of the
three members of the triplet tend to infinity. Singularities of
this kind are observed at angular momenta of unstable periodic
orbits, which do not intersect the straight line in the
$\Theta$~billiard in some orientations and do it in others.
Between singularities of this kind, every magic number except the
first and second, is the sum of two smaller ones, as proven in the
text. Note that the chosen interval of angular momentum is small
and corresponds to particles which reach the straight line close
to one of its ends.}
\end{center}
\end{figure}

As the appearance of only one or two magic numbers is restricted
to a subset of measure zero in the set of possible angular momenta
we must vary the angular momentum in order to investigate them. In
Fig.~\ref{fig_5} we show the magic numbers as a function of
angular momentum for the semicircle mushroom hat with radius $R$
and half-opening $r=R/3$. We observe discontinuities in the
behavior of the triplets of time delays (some are marked with
arrows in the figure). Precisely there one member of a magic
triplet vanishes, such that there are just two magic numbers, and
a new magic triplet emerges. The new magic number equals the sum
or difference of the other two. This apparently explains the
preliminary result that within the whole spectrum of magic numbers
each except the two smallest is the sum of two smaller ones.
\par
Moreover, we observe a singularity in Fig.~\ref{fig_5}. It is
located at an angular momentum, which is associated with a family
of marginally unstable periodic orbits~\cite{altmann}. These
periodic orbits are marginally unstable, because the smallest
change in angular momentum will cause them to rotate about the
circle center and thus eventually to hit the straight line in the
$\Theta$~billiard. The smaller the change in angular momentum, the
slower this rotation will be, and this must lead to the delay
times diverging to infinity as we approach the angular momentum of
the periodic orbit. This is clearly seen in Fig.~\ref{fig_5}. Note
that the slopes of the flanks formed by the discontinuities to the
left and the right of the angular momentum associated with the
singularity are different. Between zero angular momentum and the
first singularity as well as in each interval between two
singularities only discontinuities of the type discussed above may
happen. Therefore, within an interval, each magic number except
the two smallest ones at the minimum equals the sum of two smaller
magic numbers. However, there is no such relation between magic
numbers from two different intervals.
\begin{figure}[b]
\begin{center}
\epsfig{figure=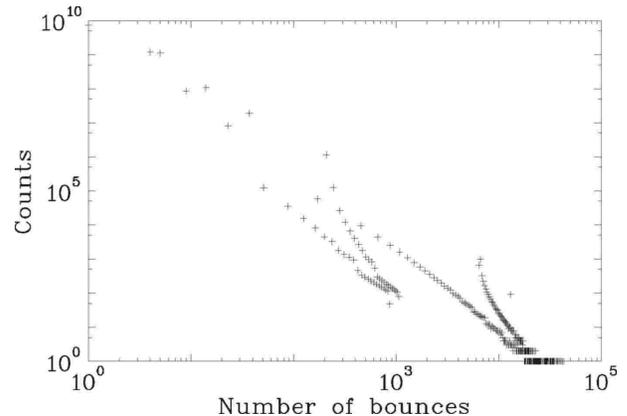,width=8cm,angle=0} \caption{\label{fig_6}
Counts of the numbers of bounce particles starting from the
straight line of half-length $r=R/3$ in the $\Theta$~billiard
experience with the circle boundary until they reach the straight
line again; they were obtained using the linear inequality
Eq.~(\ref{ineq}) for the orientation of line segments between two
reflections. Note the persisting selectivity. The organized and
regular long time behavior (large number of bounces) is dominated
by unstable periodic orbits, while the short time behavior (small
number of bounces) cannot be explained by those.}
\end{center}
\end{figure}
\par
The fact that most magic numbers can be decomposed into two
smaller ones, immediately explains why we observe a strong
selectivity in the spectra of magic numbers. In
Fig.~\ref{fig_6}~we show the counts of bounce numbers for $r=R/3$
in a double logarithmic plot up to 50000 bounces obtained using
Eq.~(\ref{ineq}). We clearly see that scarcity dominates the
picture. Note that the sequence of the magic numbers starts with
the Fibonacci like behavior as discussed above. Large magic
numbers are dominated by the marginally unstable periodic orbits,
which cause two organized sequences of magic numbers with a
certain periodicity for each singularity, one for angular momentum
values below, the other for values above the singularity.
Therefore we understand, why magic numbers may consist of subsets
which show a periodicity as observed in Ref.~\cite{altmann}. For
our purpose, Fig.~\ref{fig_6}~teaches us two facts. First, that we
are really able to solve the problem also for very large bounce
numbers. Second, that the small bounce numbers display a
selectivity, that may be accessible to experiments as the allowed
numbers typically translate into fairly distinct physical times.
\begin{figure}
\begin{center}
\epsfig{figure=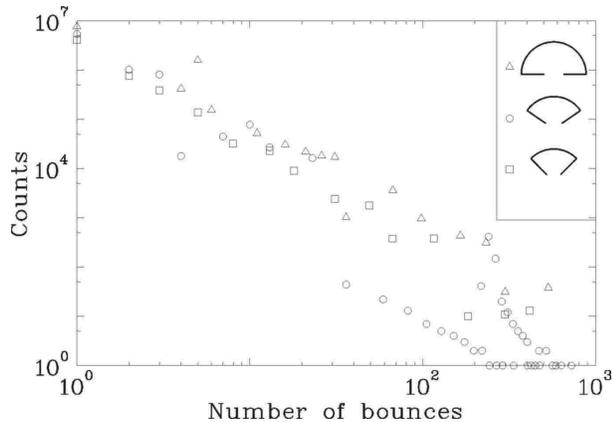,width=8cm,angle=0} \caption{\label{fig_7}
Counts of magic numbers for mushroom hats with half-opening
$r=0.3R$ and different shapes: 180$^\circ$-mushroom hat (open
triangles), $180^\circ\times (\sqrt5-1)/2$-mushroom hat (open
circles), and 90$^\circ$-mushroom hat (open squares) [see inset
and Figs.~1(a) and 1(d)]. Note that the selectivity is observed in
all examples, as the density of points remains almost constant in
the logarithmic plot. For wider openings there is no
correspondence between the number of reflections and the physical
time any more.}
\end{center}
\end{figure}
\par
It turned out that some magic numbers of a mushroom hat can
actually be calculated analytically. For every mushroom with a
central stem the first magic number equals one, $n_1=1$, as in the
limiting case of vanishing angular momentum, every orbit leaves
the mushroom hat after one reflection. Furthermore, the second
magic number of all triplets, whose first magic number is $n_1=1$
is given by
\begin{equation}
n_2=[\frac{\pi-2\chi}{\pi-2\psi}]\,,
\label{magnum1}
\end{equation}
where $[x]$ denotes the integer part of $x$. This result is
obtained by evaluating the inequality Eq.~(\ref{ineq}). It is
valid for all triplets with first magic number one, i.e., if
\[M\le\frac{1}{\sqrt{1+1/r^2}}\,.\] Using the sum rule, the third
magic number is given by $n_3=n_2+1$.
\par
For very small angular momenta, the angles $\psi(M)$ and $\chi(M)$
can be expanded in a Taylor series, leading with
Eq.~(\ref{magnum1}) to the simple expressions $n_1=1$,
$n_2=[1/r]+1$, and $n_3=[1/r]+2$ for the very first magic triplet
within a spectrum of magic numbers of an arbitrary mushroom hat.

\section{Generalizations to other mushroom billiards}
All results presented so far were given for a $180^\circ$-mushroom
billiard implying a period of $\pi$ for the line intervals. For
hats with an angle $\Omega$ smaller than $\pi$ the entire
argumentation holds as above except that the period $\pi$ is
replaced by $\Omega$. Note that this corresponds to a "generalized
$\Theta$~billiard", where the straight line rotates each time the
straight wall constituting the "underside" of the hat is hit. If
the mushroom hat has an asymmetric opening, the straight line must
be shifted even for the $180^\circ$-billiard, as the asymmetry is
reversed each time the particle hits its underside. The chain of
arguments becomes a little more tedious, but still goes through.
Thus the description of the dynamics inside the mushroom hat in
terms of the translation of orbit and line intervals is indeed
general, such that the results hold for a much wider range of
mushroom billiards. In Fig.~\ref{fig_7}~we show bounce numbers in
a double logarithmic plot for several angles and openings. The
important point is, that we see no qualitative difference in the
short time behavior: the selectivity remains in all cases.
Furthermore, the magic numbers again combine to triplets (not
shown in the figure).
\begin{figure}
\begin{center}
\epsfig{figure=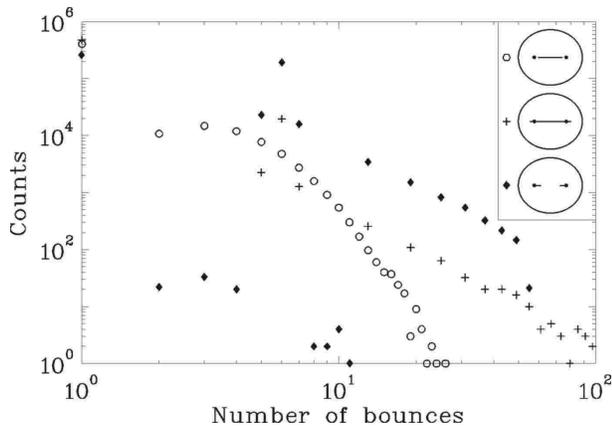,width=8cm,angle=0} \caption{\label{fig_8}
Magic numbers of an elliptic $\Theta$~billiard, corresponding to a
mushroom billiard with an elliptic hat [see Fig.~1(b)], with the
length of the semimajor axis chosen equal to $R$, that of the
semiminor axis equal to $\sqrt{3}/2R$ while the focal points have
a distance of $0.5R$ from the center. Circles refer to a straight
line chosen symmetrically between the focal points (length
$2r=0.99R$), where no selectivity is observed. For a straight line
of length $2r=1.1R$ selectivity is evident even for long times,
because here only deformed starlike periodic stable orbits of the
ellipse survive. Those particles, which intersect the straight
line between the focal points, will intersect it again after just
one bounce with the elliptic boundary. The diamonds refer to the
case of two straight lines of length $0.1R$, whose outer endpoints
are located in the focal points. They cut the caustic of the
stable orbits of particles crossing the horizontal line between
the two focal points from outside. Again a selectivity is
observed.}
\end{center}
\end{figure}
\par
We may finally go one step further and ask, what happens with
elliptic hats \cite{bunimo4}. There we must remember, that such a
hat displays two families of stable particle orbits, one
corresponding to particles travelling between the two focal
points, the other one to particles surrounding them. We performed
numerical calculations for three different choices of the
positions of the straight lines for an elliptic $\Theta$~billiard
using the reflection program. In the first case, the straight line
was positioned between the two focal points and in the second it
extends across the focal points. The results are shown in
Fig.~\ref{fig_8}. We see no selectivity in the first case
(circles) and a picture very similar to those obtained for the
semicircle hats in the second case (crosses). In the second case
particles moving on orbits, which start between the focal points
all reach the straight line again after only one bounce with the
ellipse boundary. The orbits surrounding the two focal points on
the other hand are just deformed stars. The line segments of such
stars and therefore their elliptic caustic cross the straight line
in a way similar to that of those in a circular $\Theta$-billiard.
This brought us to the assumption, that a necessary condition for
the selectivity is that the opening of the mushroom hat cuts a
limiting caustic from outside. In order to confirm this picture we
performed a third numerical simulation, where two straight lines
were positioned in the elliptic $\Theta$~billiard, which extend
from the focal points to the interior (see inset in
Fig.~\ref{fig_8}). In this case the hyperbolic caustic of those
stable orbits of particles which cross the horizontal line between
the focal points is cut by the straight line from outside, leading
indeed again to the selectivity in the magic numbers (diamonds in
Fig.~\ref{fig_8}). Note that these calculations were performed
with the simple reflection program. An analysis similar to that
for semicircular hats is also possible for elliptic hats, but
shall not be carried out here.

\section{Conclusion}
We have been able to show, that mushroom hats with openings in the
straight boundaries have a very selective behavior concerning the
number of bounces a particle has with the curved boundary before
it escapes. Indeed while the magic number one always occurs we can
to some extent design the subsequent numbers and hence the
corresponding delay times. We expect that the selectivity of the
possible delay times in the hat of the mushroom should show up not
only in classical billiards, but also in billiards excited with
microwaves. In such experiments the open hat mushroom billiard is
excited with a single pulse input signal and the output time
signal is expected to be nonperiodic, in contrast to acoustical
echoes we know from experience and scattering echoes analyzed in
some detail in Refs. \cite{mejia,carlosnjp,friedrich}. This opens
new avenues in billiard research. An experimental device for the
measurement of delay times in open mushroom billiards can be
constructed, based on the analogy between open mushroom hats and
the $\Theta$~billiard, with a flat cylindric microwave resonator
of circular shape. The straight line defining in the classical
billiard the starting and the endpoints of orbits which correspond
to particles entering and leaving the mushroom hat can be realized
by a strip of microwave absorbing material with an emitting and a
receiving antenna positioned close to the strip. To test how the
results presented in this paper carry over to wave mechanics, we
plan such a microwave experiment with a superconducting cavity
proceeding along similar lines as in the one reported in Ref.
\cite{friedrich} in order to investigate the time response. It is
though important to repeat, that the equivalence between time and
bounce numbers holds only for small openings. Once we have
starlike and nonstarlike polygon orbits coexisting the
corresponding delay times get mixed, and while the selectivity of
bounce numbers may persist, their physical importance may be
marginal in this case. Also we must be aware, that the analogy
will break down at long times, but these are probably of marginal
interest anyway.
\begin{acknowledgments}
The authors are grateful to L. Bunimovich, F. Leyvraz, and E.
Altmann for useful discussion. This work was supported by the DFG
within the Sonderforschungsbereich 634 as well as by the projects
DGAPA Contract No. IN-101603 and CONACyT Contract No. 43375-F.
Part of this work was carried out during gatherings at the Centro
Internacional de Ciencias (CIC) in Cuernavaca. One of the authors
(T.F.) received a grant from the Studienstiftung des Deutschen
Volkes during this work. Finally, this work has been completed
while one of the authors (A.R.) has been Tage Erlander Professor
2006 at Lund University supported by the Swedish Research Council.
\end{acknowledgments}

\section*{Appendix: Magic numbers for fixed angular momentum}
In this appendix we shall show that for each value of angular
momentum there exist only three magic numbers $n_l, n_r$, and
$n_s$ where the largest is always the sum of the lower ones,
$n_s=n_l+n_r$. For a given length of the straight line and a fixed
angular momentum, the angles $\psi$ and $\chi$ can be calculated
from Eq.~(\ref{psidef}) and Eq.~(\ref{chidef}), where
$0<2\psi\leq\pi$ and $0<\chi\leq\psi$ [see
Eq.~(\ref{winkelbereichchi})]. The magic numbers are determined by
using a procedure based on the inequalities Eq.~(\ref{ineq}) and
Eq.~(\ref{ineqth}) and outlined in the main text.
\par
Now consider Eqs.~(\ref{Thetan}) and (\ref{map}) which are related
to the so-called circle map \cite{slater}
\begin{equation}
\phi_{n+1}=(\phi_n+\alpha)\,\mathrm{mod}\,2\pi\,.
\end{equation}
For this map it is known for almost 40 years that angles $\phi_0$
originating in a given interval can at most have three different
recurrence times, i.e., numbers of iterations for recurrence to
the same interval \cite{slater}. The angle parameter $\alpha$
corresponds to $2\psi$ and $\phi_n$ to
$\theta_n\mathrm{mod}\,2\pi$, i.e., to the orientation of the
$n$-th line segment of the particle orbit, where the information
on the number of times the particle travelled around the circle
center until its $n$-th impact with the circle boundary is dropped
(see Eq.~(\ref{Thetan}) and the remark thereafter). Thus this map
indicates one for fixed angular momentum and we can use the
results presented in \cite{slater} to argue that, for a fixed
angular momentum, we have at most three different delay times.
\par
An intuitive proof of this result is given here based on the
translation of the orbit intervals defined in the main text after
Eq.~(\ref{ineq}). This yields an understanding of what happens
when the angular momentum varies. As an important further result
we shall see that for a fixed angular momentum less than three
delay times occur on a subset of measure zero. Since we only
consider particle orbits which start on the straight line in the
$\Theta$-billiard before their first encounter with the circular
boundary, the first orbit interval coincides with the first line
interval around zero. The first particles reach the straight line
when there is the next partial overlap of an orbit interval with a
line interval at the right hand side or at the left hand side.
Their particle orbits end there. Accordingly, this part of the
orbit interval is cut off the orbit interval in the next shift.
The remaining part evolves until the next overlap is reached.
\par
We first proceed to show in (i) that if at the first overlap after
$n_r$ reflections of the particle at the circle boundary, that is
shifts of the orbit interval by $2\psi$, the orbit interval is cut
partially from the right hand side and at the second overlap after
$n_l$ reflections from the left hand side, then the rest of the
interval overlaps with a line interval at latest after
$n_s=n_r+n_l$ reflections. The same holds in the opposite case
when the orbit interval is first cut from the left and then from
the right. Next we show in (ii) that it is not possible that the
orbit interval is shortened at the same side for the first two
subsequent overlaps. As a further step, we demonstrate in (iii)
that there is no way to find a further overlap between the first
two overlaps and the one after $n_s$ reflections. Finally we show,
that the possibility of overlapping the complete orbit interval
with less than three coincidences of the orbit and a line interval
is limitted to a subset of measure zero within the set of all
allowed angular momenta.
\par
In order to find the magic numbers it is sufficient to consider
the evolution of the orbit interval center ($oc_n$) and the line
interval center ($lc_m$) with increasing numbers of shifts
$n=0,1,2,\cdots$ and $m=0,1,2,\cdots$ of the orbit and line
intervals, respectively, where $oc_n$ is located at $n2\psi$ and
$lc_m$ at $m\pi$. The $n$-th orbit interval partially overlaps
with the $m$-th line interval, if the distance between $oc_n$ and
$lc_m$ is less than the length of the orbit and the line
intervals, $2\chi$ [see Eq.~(\ref{ineq})]. It is important to note
that the length of the orbit and the line intervals is smaller
than the distance between two subsequent $oc$'s, $2\chi\leq
2\psi$, and the distance is smaller than that of two subsequent
$lc$'s, $2\psi\leq\pi$.
\par
\noindent (i) Assume that after $n_r$ shifts of the orbit interval
and $m_1$ shifts of the line interval the distance between
$oc_{n_r}$ and $lc_{m_1}$ is smaller than $2\chi$, and that the
orbit interval there partially overlaps at its right end with the
line interval. Then $oc_{n_r}$ is located to the left of
$lc_{m_1}$ along the angle axis at a distance
$\epsilon_1=m_1\pi-n_r2\psi <2\chi$. The length of the overlapping
part of the orbit interval equals $L_1=2\chi -\epsilon_1$. Further
assume, that after $n_l$ shifts the remaining part of the orbit
interval partially overlaps at its left end with the $m_2$-th
shift of the line interval. Then $oc_{n_l}$ is located to the
right of $lc_{m_2}$ at a distance $\epsilon_2=n_l2\psi-m_2\pi
<2\chi$ and the length of the overlapping part equals $L_2=2\chi
-\epsilon_2$. The length of the remaining part of the orbit
interval equals
\begin{eqnarray}
L &=&2\chi-L_1-L_2\\
  &=&\epsilon_1+\epsilon_2-2\chi\\
  &=&(n_l-n_r)2\psi-(m_2-m_1)\pi -2\chi.
\label{Leq}
\end{eqnarray}
If $L=0$, then already after two overlaps of an orbit with a line
interval, all particles will have returned to the straight line,
such that there are only two magic numbers; this case will be
treated in part (iv) of the proof. The case $L<0$, where more than
just the remaining part of the orbit interval overlaps with the
line interval, may be excluded, as this would imply that there is
another magic number $n^*=(n_l-n_r)<n_l$ and an
$m^*=(m_2-m_1)<m_2$, where the distance between $oc_{n^*}$ and
$lc_{m^*}$ is less than $2\chi$ [see Eq.~(\ref{Leq})], in
contradiction to the assumption. For $L>0$, the remaining part is
adjacent to the line interval at its left end; its left endpoint
is located at a distance $\chi$, its right endpoint at a distance
$\epsilon_1 +\epsilon_2 -\chi$ from $lc_{m_2}$. Exactly as after
the first $n_r$ and $m_1$ shifts of the orbit and the line
interval, after $n_r$ and $m_1$ further shifts of the intervals,
the location of $oc_{n_s}$ ($n_s=n_r+n_l$), and therefore that of
the remaining part, will have moved by a total amount of
$\epsilon_1$ to the left with respect to $lc_{m_3}$
($m_3=m_1+m_2$) as compared to their initial relative locations,
i.e., that of $oc_{n_l}$ and $lc_{m_2}$. Accordingly, $oc_{n_s}$
is located at a distance $\vert\epsilon_2-\epsilon_1\vert$ from
$lc_{m_3}$, which is smaller than $2\chi$. The left endpoint of
the part of the orbit interval that remained after $n_l$ shifts,
is then located at a distance $\vert\chi -\epsilon_1\vert$, the
right endpoint at a distance $\vert\epsilon_2-\chi\vert$ from
$lc_{m_3}$. As both distances are smaller than $\chi$, this part
completely overlaps with the line interval. Therefore, provided
the orbit interval is first cut at its right, then at its left end
at the magic numbers $n_r$ and $n_l$, the remaining particles all
will have reached the straight line again latest after
$n_s=n_r+n_l$ shifts of the orbit interval, or equivalently,
bounces of the particle with the circle boundary. The same
argumentation holds if the interval is first cut from the left and
then from the right hand side.
\par
If the first magic number equals 1, as in the example shown in
Fig.~\ref{fig_4}, we already arrived at the end of the proof. In
this case, the orbit interval will be cut at its right end after
one shift of the line and the orbit interval and $oc_1$ will be
located at a distance $\epsilon_1=\pi-2\psi$ to the left of
$lc_1$. At each shift $n$ and $m=n$ of the intervals, the distance
of $oc_n$ from $lc_m$, which is located to its right, will
increase by an amount $\epsilon_1$ until the orbit interval
overlaps with the line interval to its left after $n_l$ shifts,
i.e., until the distance between $oc_{n_l}$ and $lc_{n_l-1}$ is
smaller than $2\chi$. Hence, at the second overlap the orbit
interval will be cut at its left end, at a magic number $n_l$.
Then, as shown in part (i) of the proof, at $n_s=n_l+1$ all
particles will have reached the straight line, and we obtain the
magic triplet $1,n_l,n_l+1$. The explicit values can be computed
with Eq.~(\ref{magnum1}).
\par
\noindent (ii) Assume, that the first magic number is larger than
1 and that the orbit interval overlaps at its left hand side after
$n_{l1}$ and again after $n_{l2}$ shifts of oc with respectively
the $m_1$-th and the $m_2$-th line interval. Then, $oc_{n_{l1}}$
and $oc_{n_{l2}}$ will be located to the right of $lc_{m_1}$ and
$lc_{m_2}$, respectively, at a distance
\begin{equation}
\epsilon_1=n_{l1}2\psi-m_1\pi<2\chi
\end{equation}
for the first and
\begin{equation}
\epsilon_2=n_{l2}2\psi-m_2\pi<2\chi
\end{equation}
for the second overlap. The length of the first overlapping part
equals $L_1=2\chi-\epsilon_1$, that of the second
$L_2=2\chi-\epsilon_2$, and the orbit interval will be further
shortened at the second overlap only, if $L_2$ is larger than
$L_1$, that is, if $\epsilon_2<\epsilon_1$. Then
\begin{equation}
0<\epsilon_1-\epsilon_2=(m_2-m_1)\pi-(n_{l2}-n_{l1})2\psi<2\chi\,.
\end{equation}
This implies, that there is another magic number
$n^*=n_{l2}-n_{l1}<n_{l2}$ and a number $m^*=m_2-m_1$, where the
distance $\epsilon_1-\epsilon_2$ of $oc_{n^*}$ from $lc_{m^*}$ is
less than $2\chi$ (right hand side of the inequality) and the
orbit interval is shortened at its right hand side (left hand side
of the inequality). This is in contradiction to the assumption.
\par
\noindent (iii) Let the orbit interval be shortened at the right
hand side after $n_r$ reflections and at the left hand side after
$n_l$ reflections. Assume, that the third magic number is smaller
than the sum of the first and second, $n^*<n_r+n_l$. If after
$n^*$ shifts the orbit interval is shortened again at the left
hand side, then, according to (ii), there exists an additional
magic number $n^{**}=n^*-n_l<n_r$, where the orbit interval is
shortened at the right hand side, in contradiction to the
assumption. If on the other hand $n^*$ shortens the orbit interval
at the right hand side, then, by proceeding as in (ii), it can be
shown that there must be another magic number $n^{**}=n^*-n_r<n_l$
where the orbit interval overlaps at the left hand side with a
line interval, again in contradiction to the assumption.
\par
\noindent (iv) Assume that the orbit interval after $n_r$ shifts
partially overlaps  at its right hand side with a line interval,
and after $n_l$ shifts the remaining part completely overlaps with
a line interval. This situation corresponds to the case $L=0$ in
part (i), i.e., if $(n_l-n_r)2\psi-(m_2-m_1)\pi=2\chi$, then there
are only two magic numbers. For a given length of the straight
line, this equality is fulfilled only for a set of measure zero
from the whole set of allowed angular momenta [see
Eq.~\ref{psidef}) and Eq.~(\ref{chidef})]. As we vary the angular
momentum for a fixed length of the straight line, the angles
$\psi$ and $\chi$ change continuously, and a third magic number
emerges because the first and second overlap no longer cover the
whole orbit interval. There, we observe a discontinuity in
Fig.~\ref{fig_5}.
\par
The arguments given in this appendix only hold for finite magic
numbers. In the case of the parabolic manifolds described in Ref.
\cite{altmann} only one magic number is left finite. The values of
angular momentum corresponding to those orbits are also of measure
zero in the whole set of possible angular momenta. In
Fig.~\ref{fig_5}, we observe a singularity at these values of
angular momentum.
\par
Finally we should note, that the proof also holds, if the line
interval is shifted by an arbitrary angle $2\chi<\Omega<\pi$, as
we only used the fact, that the length of the orbit and the line
interval is smaller than the distance between two subsequent line
intervals or orbit intervals.

\end{document}